\documentclass[twocolumn,longbibliography,aps,prl,superscriptaddress,showpacs,floatfix]{revtex4-1}
\usepackage{graphics,epsfig,graphicx}
\usepackage{amsbsy,gensymb}
\usepackage{amsmath}
\usepackage{bm}
\usepackage{amsfonts}
\usepackage{cancel}
\usepackage{multirow}
\begin{document}

\title{Saturation properties of helium drops from a Leading Order description}

\author{A. Kievsky} 
\affiliation{Istituto Nazionale di Fisica Nucleare, Largo Pontecorvo 3, 56100 Pisa, Italy}
\author{A. Polls} 
\author{B. Juli\'a D\'\i az}
\affiliation{Departament de F\'\i sica Qu\`antica i Astrof\'\i sica, Facultat de F\'\i sica,
Universitat de Barcelona, E-08028 Barcelona, SPain}
\author{N. K. Timofeyuk}
\affiliation{Department of Physics, University of Surrey, Guildford, Surrey GU2 7XH, United Kingdom}

\begin{abstract}
Saturation properties are directly linked to the short-range scale of the two-body
interaction of the particles. The case of helium is particular, from one hand the 
two-body potential has a strong repulsion at short distances. On the other hand, 
the extremely weak binding of the helium dimer locates this system very close 
to the unitary limit allowing for a description based on an effective theory. At 
leading order of this theory a two- and a three-body term appear, each one 
characterized by a low energy constant. In a potential model this description 
corresponds to a soft potential model with a two-body term purely attractive plus a 
three-body term purely repulsive constructed to describe the dimer and trimer 
binding energies. Here we analyse the capability of this model to describe the 
saturation properties making a direct link between the low energy scale and the 
short-range correlations. 
We will show that the energy per particle, $E_N/N$, can 
be obtained with reasonable accuracy at leading order extending the validity of
this approximation, characterizing universal behavior in few-boson systems close to 
the unitary limit, to the many-body system.
\end{abstract}

\maketitle

{\sl \noindent Introduction.}
At the beginning of the eighties strong efforts were done to calculate ground 
state properties of $^4$He and $^3$He droplets containing specific number 
$N$ of atoms~\cite{kalos,usmani,pandha1,pandha2}. After computing the energy 
per particle, $E_N/N$, and the rms radii of the droplets it was possible to 
study the evolution of these quantities as $N\rightarrow\infty$. For example, 
a liquid-drop formula was proposed to fit $E_N/N$ in terms of $x=N^{-1/3}$
\begin{equation}
E_N/N=E_v + E_s x + E_c x^2
\label{eq:ldf}
\end{equation}
with $E_v$, $E_s$ and $E_c$, the volume, surface and curvature terms respectively.
A similar behavior, in powers of $x$, has been proposed for the unit radius, 
defined in terms of the rms radius $\langle r^2\rangle^{1/2}$, as 
$r_0(N)=\sqrt{5/3}\, \langle r^2\rangle^{1/2} N^{-1/3}$. Extrapolated results for the 
infinite liquid were obtained from calculations on droplets using different
values of $N$. The motivations for that study were twofold, from one side the 
theoretical results obtained with realistic interatomic potentials could be compared 
to experimental results. To this respect the calculation on the infinite system, 
liquid $^4$He at equilibrium density, predicts a value $E_v=-7.11$ K using the high 
quality potential HFDHE2 from Aziz et al.~\cite{aziz1}, in very good agreement with 
the experimental value of $-7.14$ K at a density of 0.0219 \AA$^{-3}$. This can be 
seen as a successful application of the potential theory to describe ground state 
properties of liquid helium. A second motivation was to analyze the capability of 
the extrapolation formulas to predict the properties of the infinite system 
using results computed in droplets having at most a few hundred atoms. It 
was shown that stable values of $E_v$ and the surface tension $t=E_s/4\pi r_0^2(\infty)$ 
could be obtained in agreement with those calculated in 
the infinite system. This analysis gave support to the liquid-drop formulas used 
in nuclear physics to predict nuclear matter properties. To be noticed that 
whereas different properties can be measured in infinite liquid helium this is not 
the case for infinite nuclear matter. 

Droplets of bosonic helium attracted attention in the nineties due to the fact that the dimer 
composed by two $^4$He atoms is very loosely bound. Its energy is $E_2\approx 1$ mK,
and the two-body scattering length, $a\approx 100\;$\AA , has a very large value 
if compared to the typical length of the system, the van der Waals length $\ell_{vdw}$, 
which for two helium atoms is $\ell_{vdw}\approx 2.5\;$\AA. When $a \gg \ell_{vdw}$ 
the system can be studied in first approximation in the zero-range 
limit. It provides a good approximation for 
shallow states in which the particles stay most of the time outside the interaction 
region and, accordingly, the low energy dynamics does not depend on the details 
of the interaction. 
Moreover  $E_2 \approx \hbar^2 /(ma^2)$, with $m$ the boson 
mass, vanishes at the unitary limit, corresponding to $a\to\infty$. 
As demonstrated by Efimov in a series of papers~\cite{efimov1,efimov2}, 
the three-body system has a geometrical series of excited states 
that accumulate at zero energy. This is called the Efimov effect and was experimentally 
confirmed more than three decades after its prediction~\cite{kraemer2006}. 

At present days there is an intense experimental activity~\cite{ferlaino2011,machtey2012,roy2013,dyke2013}
dedicated to study the behavior of few-body systems close to the unitary limit. To 
this respect the helium trimer was indicated as a candidate for a direct observation
of an Efimov type excited state. The possibility of observing Efimov states in small 
clusters of helium has triggered an intense experimental activity using ultracold 
jets of helium going through a diffraction grating~\cite{toennies}. Though it was 
not possible to extract specific energy values, the diffraction patterns were used to 
identify the number of atoms in the droplets. This research culminated recently 
with a measurement of the ground and excited state of the helium trimer giving a 
direct confirmation of the existence of Efimov states~\cite{doener}.

Helium drops have been studied 
using modern helium-helium interactions~\cite{rafa,boronat}. In particular in Ref.~\cite{lewerenz} a 
diffusion Monte Carlo (DMC) method has been used to study clusters up to 10 atoms 
interacting through the Tang, Toennies and Yiu (TTY)
potential~\cite{TTY}. From a more general perspective,
trimers and tetramers have been studied with different interactions in which the 
potential strength has been varied in order to drive the system to the unitary 
limit~\cite{greene,barletta2001,hiyama1,hiyama2}. When a two-boson system interacting 
via a short-range potential is close to the unitary limit, the three-boson system 
shows universal behavior. Its spectrum is governed by the two-body scattering 
length $a$ and the three-body parameter $\kappa_*$ defines the energy of the $n_*$ 
level at the unitary limit, $\hbar^2\kappa_*^2/m$. 
The system manifests a discrete scale invariance (DSI), the ratio of 
binding energies for two consecutive states is $E^n_3/E^{n+1}_3=e^{2\pi/s_0}$, with the 
universal number $s_0\approx 1.00624$~\cite{report}. 
The studies using potential models have shown that this 
description is very well fulfilled if range corrections are taken into account~\cite{raquel}.

A three-boson system close to the unitary limit can be described using an effective field theory 
(EFT)~\cite{bedaque99,bedaque2000}. At leading order (LO) the effective Hamiltonian 
includes a two-body and a three-body contact term. The strength of the two terms determine
the values of $a$ and $\kappa_*$. This kind of studies have triggered the idea of describing
the dimer and trimer using a soft potential model consisting in a two- plus a three-body 
term in which the strengths can be fixed to describe some particular observables, 
for example the dimer and trimer binding energies. This Hamiltonian can be used to 
solve the Schr\"odinger equation for systems with $N>3$ and the agreement (or differences) 
obtained from comparisons to experimental data or results obtained with more realistic 
interactions can be analysed. This strategy has been explored in 
Refs.~\cite{kievsky2011,gatto2011,gatto2012} in which the ground state energy of small clusters 
of helium calculated using a soft potential model
results extremely close (within a few percent) to that one obtained using a realistic 
helium-helium interaction.

From the above discussion we observe two, very distinctive, 
descriptions of light helium clusters. On one hand, strong efforts have been done
to determine the best possible helium-helium interaction. Different models exist 
in the literature and they have been tested in drops as well as in infinite liquid. 
On the other hand the large scattering length of the helium-helium system indicates 
that the helium trimer and tetramer show universal behavior. The particular form 
of the potential is not important and many features can be determined from a few 
experimental data, such as $a$ and the trimer ground state energy $E_3^0$ (or first excited state $E_3^1$). 
Accordingly a soft potential model can be constructed in order to reproduce those 
observables. 
Here we want to determine saturation properties 
of the infinite system from calculations on helium drops described 
using a soft potential model making a direct link between 
the low energy scale (or long-range correlations) 
and the high energy energy scale (or short-range correlations). Moreover this analysis 
will clarify whether a four-body force is needed at a LO description. 

In order to treat the helium clusters with increasing number of particles we 
use two different methods. We expand the many-body wave function 
in the hyperspherical harmonic (HH) basis and calculate the ground state energy 
for increasing values of the grand orbital quantum number $K$. The method using 
two- and three-body potentials is described in Refs.~\cite{timofeyuk1,timofeyuk2}. 
Depending on the range of the three-body force the pattern of convergence in 
terms of $K$ could not be sufficiently fast to guarantee a converged value for 
the energy. In order to decrease the uncertainty in the energy of the system we 
use a DMC code to calculate the ground state of the system. 
From the combination of the two methods we obtain converged values for the 
ground state energy.

\begin{figure}[t]
\includegraphics[scale=0.4,angle=-90]{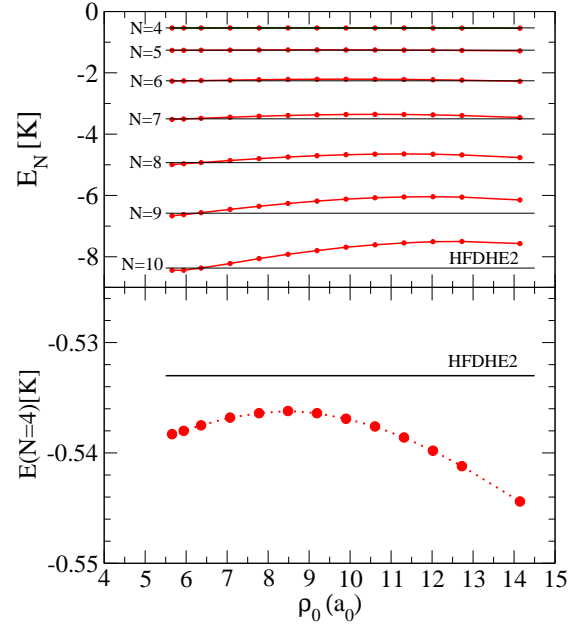}
\caption{Color online. Binding energies using the SGP for different values of the 
three-body force range $\rho_0$ (red dots) at the indicated $N$ values (upper panel). 
The specific case of $N=4$ is shown in the lower panel.
As a reference, the values of the HFDHE2 
potential are also shown (black solid lines).}
\label{fig1}
\end{figure}

\begin{figure}[t]
\includegraphics[scale=0.35,angle=-90]{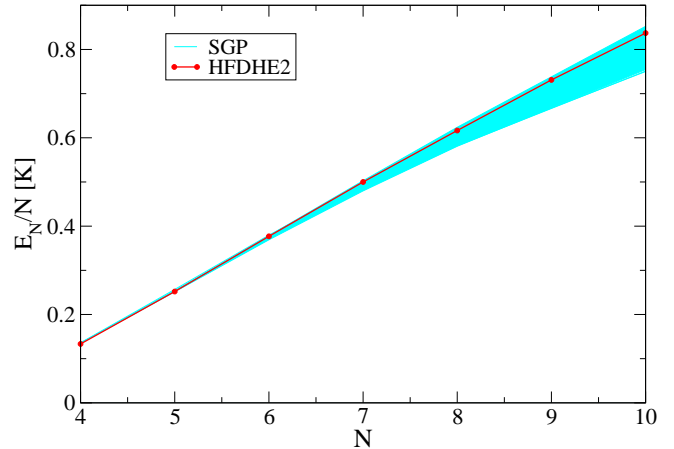}
\caption{Color online. Binding energy per particle as a function of the number 
of particles $N$. The results of the SGP for different values of the three-body 
force range $\rho_0$ are shown as the cyan band. As a reference, the values of 
the HFDHE2 potential form Ref.~\cite{pandha1} are shown as solid (red) circles.}
\label{fig2}
\end{figure}

{\sl \noindent The potential model.}
To study the ground state energy of the $N$ boson system we use as a reference calculations
on helium drops interacting through the HFDHE2 potential. Results using the Green's function Monte 
Carlo (GFMC) method are available as well as results using a Variational Monte 
Carlo (VMC) approach. The soft potential model is constructed using a gaussian 
representation of the HFDHE2 potential as
\begin{equation}
 V(r_{ij})= V_0 e^{-r_{ij}^2/r_0^2}
\end{equation}
in which the two parameters, $V_0$ and $r_0$, are determined from the dimer 
energy, $E_2=0.83012\,$mK, and the two-body scattering length $a=235.547\,a_0$,
with $a_0$ the Bohr radius. 
These quantities are described with good accuracy using $V_0=1.208018\,$K 
and $r_0=10.0485\,a_0$ (with $\hbar^2/m=43.281307\,$K$a_0^2$). Using this
potential the binding energy of the trimer ground state is $139.8\,$mK, 
this value is greater than the value obtained with the HFDHE2 potential of 
$117.3\,$mK. Accordingly the two-body soft potential has to be supplemented 
with a slightly repulsive three-body force. This well known characteristic 
corresponds, in terms of EFT, to a LO description. Following 
Refs.~\cite{kievsky2011,gatto2011,gatto2012,timofeyuk1,timofeyuk2} we introduce 
a three-body force depending on the relative distances of three particles
\begin{equation}
 W(\rho_{ijk})= W_0 e^{-2\rho_{ijk}^2/\rho_0^2}
\end{equation}
where $\rho^2_{ijk}=(2/3)(r^2_{ij}+r^2_{jk}+r^2_{ki})$ and the strength $W_0$ and range 
$\rho_0$ are parameters to be fixed in order to have a reasonable description of 
the light clusters ground state binding energies $E_N$. In the following we employ
the soft-gaussian potential (SGP) model consisting on a two-body plus a three-body term. 
The SGP ground state binding energies up to $N=10$ are shown in Fig.~\ref{fig1} 
(red dots) as a function of the three-body range parameter $\rho_0$. In each case 
the strength $W_0$ is fixed to reproduce the trimer ground state of the HFDHE2 
potential ($117.3\;$mK). The SGP results are compared to those of the HFDHE2 
potential~\cite{pandha1} given in the figure as the (black) solid lines. As can be 
seen from the figure there is a slight dependence on the range $\rho_0$, with low 
values giving a better description. To show the sensitivity to the range of the 
three-body force and to analyse the behavior of the energy per particle $E_N/N$, 
in Fig.~\ref{fig2} we show this quantity as a function of $N$. We can observe 
that, for the values of $N$ given in the figure, $E_N/N$ calculated with the HFDHE2 
interaction has an almost linear behavior. The results of the SGP follow this 
tendency though a spread depending on $\rho_0$ appears as $N$ increases. 

In the present study the strength and range of the two-body gaussian potential are 
determined from $E_2$ and $a$. In a more general perspective a gaussian potential can 
be thought of as regularized contact interaction and the observables in the different 
$N$-body sectors can be studied in terms of the range of the gaussian defined as 
the inverse of the cutoff $r_0=\Lambda^{-1}$ (for a recent discussion see Ref.~\cite{bazak}). 
In this context the range of the two- and three-body forces are related. Here we 
follow a different strategy in which the two-body potential is fixed by two data 
in the $N=2$ sector. The strength of the three-body potential is determined by 
$E_3$ for different values of its range $\rho_0$. In this way the evolution 
of $E_N/N$ can be studied as a function of the parameter $\rho_0$. To be noticed 
that the two- and three-body potential terms evolve differently with $N$ since 
one is proportional to the number of pairs and the other to the number of triplets. 
The intention of using $\rho_0$ as an independent parameter is to keep the evolution 
of these two terms as close as possible to the results of the original potential. 
Eventually a particular value of $\rho_0$ can be detected as the optimum value 
to use in the description of the saturation properties of the infinite system. 

\begin{figure}[t]
\includegraphics[scale=0.50,angle=-90]{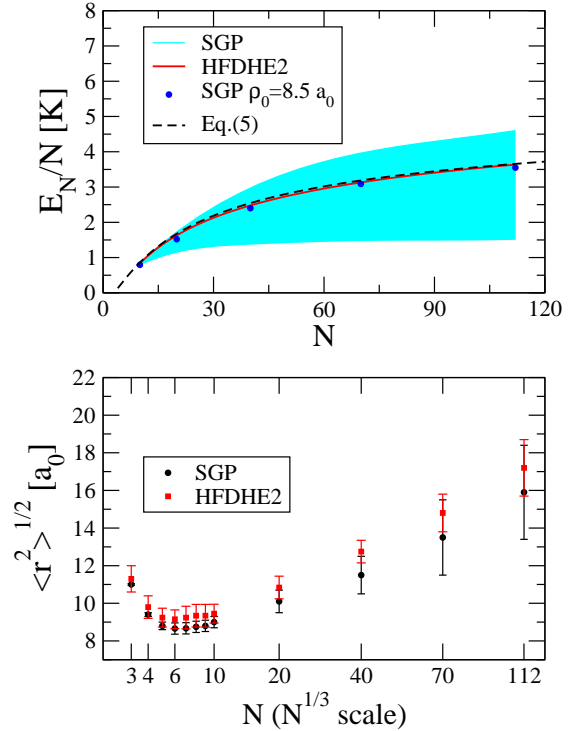}
\caption{Color online. Binding energy per particle (upper panel) and rms radii 
(lower panel) as a function of the number of particles $N$. The different values 
of the range $\rho_0$ of the SGP are shown as the cyan band (for $E/N$) or as 
error bars (rms radii) The dashed line is the prediction of Eq.(\ref{eq:onep}) (see text). 
The values of the HFDHE2 potential are also presented.}
\label{fig3}
\end{figure}

{\sl \noindent $E_N/N$ using a soft potential model.}
Here we extend the study of $E_N/N$ for increasing values of $N$. 
The calculations of Ref.~\cite{pandha1} using the HFDHE2 potential show that 
this quantity has an almost linear behavior for $N \le 10$, as discussed 
before. As $N$ is increased further $E_N/N$ 
saturates following the trend given by Eq.~(\ref{eq:ldf}). This behavior is 
confirmed by the rms radius which increases almost linearly with $N^{1/3}$ 
for $N>20$, resembling a liquid drop. Now we want to analyse the evolution of 
the binding energy using the SGP. To this aim, we calculate 
$E_N/N$ and radii up to $N=112$, this value seems to be sufficient
to determine $E_v$ from Eq~.(\ref{eq:ldf}). The results are given in Fig.~\ref{fig3}.
There is a large spread in both quantities depending on the three-body range 
$\rho_0$ given as the cyan band for $E_N/N$ and as error bars for the rms radii. 
The HFDHE2 results are inside the energy per particle band therefore an optimum 
value of $\rho_0$ can be identified. From inspection of the results this particular 
value is $\rho_0\approx 8.5\; a_0$ and corresponds to the range needed to get the 
closest value to the exact tetramer binding energy, as can be seen in the lower panel
of Fig.~\ref{fig1}. 
Using this value of $\rho_0$  it is possible to determine $E_v$, $E_s$ and $E_c$ defined 
in Eq.~(\ref{eq:ldf}). 
From the results of the SGP in the range $20\le N \le 112$ we obtain (in K)
\begin{equation}
E_N/N= 6.98 - 18.6\, x + 10.3\, x^2
\label{eq:ldf1}
\end{equation}
to be compared to the values (in K) $E_v=7.02$, $E_s=-18.8$ and $E_c=11.2$ 
and $E_v=6.91$, $E_s=-18.9$ and $E_c=12.0$ obtained with the
GFMC and VMC methods, respectively using the HFDHE2 interaction. 

The infinite unit radius $r_0(\infty)$ can be obtained from a second order expansion 
in terms of $x=N^{-1/3}$. The SGP results for the optimum $\rho_0$ value predict 
$r_0(\infty)=2.24\;$\AA, close to the GFMC result for the HFDHE2 interaction of $2.22\;$\AA
and a surface tension $t=E_s/4\pi r_0^2(\infty)$ of $0.29\;$K\AA$^{-2}$, 
close to the experimental value of $0.27\;$K\AA$^{-2}$ and the HFDHE2 GFMC result of 
$0.28\;$K\AA$^{-2}$. We consider the capability of the SGP of following 
the energy per particle and the unit radius (giving a reasonable prediction 
of the surface tension) a consequence of the propagation of the universal behavior 
observed in the three-body sector to the infinite system. This is an unexpected result.
Accordingly we can think in a different expansion of $E_N/N$ in terms of $N$
incorporating explicitly the energy values of the light droplets. Considering that
$E_3/3$ is almost negligible compared to $E_N/N$ as $N\rightarrow\infty$, we can propose 
the following formula
\begin{equation}
\frac{E_N}{N}=E^{(0)}_v\frac{1-(3/N)^{1/4}}{1+\frac{3E_4}{4E_3}(3/N)} \,\, ,
\label{eq:onep}
\end{equation}
where the exponent of $1/4$ in the numerator and the energy coefficient in the denomitar
are optimal choices to describe the GFMC results.
Using Eq.(\ref{eq:onep}) to fit the GFMC results in the region $4\le N\le 112$ the value
$E^{(0)}_v\approx 6.8$ is obtained with a comparable overall accuracy
to Eq.(\ref{eq:ldf}) as shown in Fig.~\ref{fig3} by the dashed line. If the
range of the fit is limited to the region $4\le N\le 10$, where the energy per particle increases
almost linearly, the value $E^{(0)}_v\approx 6.5$ is obtained.
A characteristic of Eq.(\ref{eq:onep}) is that  $E^{(0)}_v$
can be determined using a single value of $E_N/N$.
Making explicit the N$=4$ case we obtain
\begin{equation}
\frac{E^{(0)}_v}{E_4}=3.602\left(1+\frac{9E_4}{16E_3}\right) \; .
\label{eq:v4}
\end{equation}
This relation gives the saturation energy in units of $E_4$. Using the GFMC ratio
$E_4/E_3=4.55$ we obtain ${E^{(0)}_v}/{E_4}=12.8$. 
From this analysis it could be thought that, besides range corrections
(to evaluate in a forthcoming analysis), the saturation energy of the droplets
could be proportional to $E_4$ as $E^{(0)}_v = \xi_4 E_4$ with $\xi_4$ approaching 
a universal number at unitary in a similar way in which is defined the Bertsch parameter 
in the case of a Fermi gas~\cite{endres}.

{\sl \noindent Conclusions.}
There are two distinct approaches to describe bosonic helium drops. It is possible to 
use a realistic atomic interaction obtained from a detailed description of the electronic 
cloud. These potentials are able to describe many observables in the low and high 
energy domains, as well as transport properties. 
A different view which puts in evidence the fact that the helium system is close
to the unitary limit, is to construct a very simple potential model able to reproduce 
a few data as the dimer and trimer energies and the large value of the two-body 
scattering length. This model is constructed as a sum of a two-body (attractive) and 
a three-body (repulsive) soft terms. It can describe with good approximation 
properties that emerge as quasi universal, as for example the ratio $E_3^0/E_3^1$ between 
the ground and excited states of the helium trimer or the ratios $E_4^0/E_3^0$ and $E_4^1/E_3^0$ 
between the ground state trimer and the two levels of the tetramer~\cite{kievsky2015}. 
Our main conclusion is that
the universal properties observed in light drops propagate with the number of particles allowing
an estimate of the saturation energy from the energy of very light drops. The limiting
case is given by Eq.(\ref{eq:onep}) in which the saturation energy can be determined by the
ratio $E_4/E_3$ and one of the two values. Following some ideas discussed in the 
literature~\cite{bazak,blume},
we have speculated about the universal characteristic of the ratio $E_v^{(0)}/E_4$ at unitarity.

A second observation of the present work is that a four-body interaction is not
needed to describe the saturation properties at LO. 
We can conclude that the soft-gaussian potential captures the physics of the system close to unitarity building a 
bridge between few- and many-body physics.

\begin{acknowledgements}
{\sl \noindent Acknowledgements.}
A. P. and B. J-D. acknowledge fruitful discussions with A. Sarsa and 
J. Boronat. This work is partially supported by Grant No. FIS2014-54672-P from MICINN (Spain), Grant No.
2014SGR-401 from Generalitat de Catalunya (Spain), and  
the United Kingdom Science and Technology Facilities
Council (STFC) under Grant No. ST/L005743/1.
B. J-D. is supported by a Ramon y Cajal (Talent Retingut) contract. 
\end{acknowledgements}


\begin{thebibliography}{10}
\bibitem{kalos}M.H. Kalos, M.A. Lee, P.A. Whitlock, and G.V. Chester, Phys. Rev. B {\bf 24}, 115 (1981).
\bibitem{usmani} Q. N. Usmani, S. Fantoni, and V.R. Pandharipande, Phys. Rev. B {\bf 26}, 6123 (1982).
\bibitem{pandha1} V. R. Pandharipande, J.G. Zabolitzky, S.C. Pieper, R.B. Wiringa, and U. Helmbrecht,
        Phys. Rev. Lett. {\bf 50}, 1676 (1983).
\bibitem{pandha2} V. R. Pandharipande, S.C. Pieper, and R.B. Wiringa,
        Phys. Rev. B {\bf 34}, 4571 (1986).
\bibitem{aziz1}R.A. Aziz, V.P.S. Nain, J.S. Carley, W.L. Taylor, and G.T. McConville,
        J. Chem. Phys. {\bf 70}, 4330 (1979).
\bibitem{efimov1} V. Efimov, Phys. Lett. B {\bf 33}, 563 (1970).
\bibitem{efimov2} V. Efimov, Sov.J. Nucl. Phys. {\bf 12}, 589 (1971).
\bibitem{kraemer2006} T. Kraemer et al., Nature {\bf 440}, 315 (2006).
\bibitem{ferlaino2011}F. Ferlaino, A. Zenesini, M. Berninger, B. Huang,
H.C. N\" agerl, and R. Grimm, Few-Body Syst. \textbf{51}, 113 (2011).
\bibitem{machtey2012} O. Machtey, Z. Shotan, N. Gross, and L. Khaykovich,
Phys. Rev. Lett. {\bf 108}, 210406 (2012).
\bibitem{roy2013} S. Roy, M. Landini, A. Trenkwalder, G. Semeghini, G. Spagnolli,
A. Simoni, M. Fattori, M. Inguscio, and G. Modugno,
Phys. Rev. Lett. {\bf 111}, 053202 (2013).
\bibitem{dyke2013} P. Dyke, S.E. Pollack, and R.G. Hulet, Phys. Rev. A {\bf 88},
023625 (2013).
\bibitem{toennies}R.E. Grisenti, W. Sch\"ollkopf, J.P. Toennies, J.R. Manson, T.A. Savas, and Henry I. Smith,
        Phys. Rev. A {\bf 61}, 033608 (2000).
\bibitem{doener} M. Kunitski et al., Science {\bf 348}, 551 (2015).
\bibitem{rafa} R. Br\"uhl, R. Guardiola, A. Kalinin, O. Kornilov, J. Navarro, T. Savas, 
and J. P. Toennies, Phys. Rev. Lett. {\bf 92}, 185301 (2004). 
\bibitem{boronat}P. Stipanovi\'c, L.V. Marki\'c, and J. Boronat, 
J. Phys. B: At. Mol. Opt. Phys. {\bf 49}, 185101 (2016).
\bibitem{lewerenz} M. Lewerenz, J. Chem. Phys. {\bf 106}, 4596 (1997).
\bibitem{TTY} K.T. Tang, J. P. Toennies, and C.L. Yiu, Phys. Rev Lett. {\bf 74}, 1546 (1995).
\bibitem{greene}B.D. Esry, C.D. Lin, and Chirs H. Greene, Phys. Rev. {\bf A 54}, 394 (1996).
\bibitem{barletta2001} P. Barletta and A. Kievsky, Phys. Rev. A {\bf 64}, 042514 (2001).
\bibitem{hiyama1} E. Hiyama and M. Kamimura, Phys. Rev. A {\bf 85}, 062505 (2012).
\bibitem{hiyama2} E. Hiyama and M. Kamimura, Phys. Rev. A {\bf 90}, 052514 (2014).
\bibitem{report} E. Braaten and H.-W. Hammer, Phys. Rep. {\bf 428}, 259 (2006).
\bibitem{raquel} R. \'Alvarez-Rodr\'\i guez, A. Deltuva, M. Gattobigio and A. Kievsky,
Phys. Rev. A {\bf 93}, 062701 (2016).
\bibitem{bedaque99} P.F Bedaque, H.-W. Hammer, and U. van Kolck, Phys. Rev.
Lett.{\bf 82}, 463 (1999).
\bibitem{bedaque2000} P. Bedaque, H.-W. Hammer, and U. van Kolck, Nucl. Phys.
A{\bf 676}, 357 (2000).
\bibitem{kievsky2011} A. Kievsky, E. Garrido, C. Romero-Redondo and P. Barletta,
       Few-Body Syst. {\bf 51}, 259 (2011).
\bibitem{gatto2011} M. Gattobigio, A. Kievsky and M. Viviani, Phys. Rev. A {\bf 84}, 052503 (2011).
\bibitem{gatto2012} M. Gattobigio, A. Kievsky and M. Viviani, Phys. Rev. A {\bf 86}, 042513 (2012).
\bibitem{timofeyuk1} N.K. Timofeyuk, Phys. Rev. {\bf A86}, 032507 (2012).
\bibitem{timofeyuk2} N.K. Timofeyuk, Phys. Rev. {\bf A91}, 042513 (2015).
\bibitem{bazak} B. Bazak, M. Eliyahu, and U. van Kolck, Phys. Rev. {\bf A94}, 052502 (2016).
\bibitem{endres} M.G. Endres, D.B. Kaplan, J.W. Lee, and A.N. Nicholson, Phys. Rev. {\bf A87}, 023615 (2013).
\bibitem{kievsky2015}A. Kievsky and M. Gattobigio, Phys. Rev. {\bf A92}, 062715 (2015).
\bibitem{blume} D. Blume, Physics {\bf 3}, 74 (2010).



\end{thebibliography}
\end{document}